\pdfoutput=1

\documentclass[conference]{IEEEtran}
\usepackage{graphicx}
\usepackage[cmex10]{amsmath}
\usepackage{listings}
\begin{document}

\title{Defining Big Data Analytics Benchmarks for Next Generation Supercomputers}

\author{
  \IEEEauthorblockN{%
    Drew Schmidt\IEEEauthorrefmark{1},
    Junqi Yin\IEEEauthorrefmark{2},
    Michael Matheson\IEEEauthorrefmark{3},
    Bronson Messer\IEEEauthorrefmark{4}, and
    Mallikarjun Shankar\IEEEauthorrefmark{5}
  }
  \IEEEauthorblockA{%
    Oak Ridge National Laboratory,
    Oak Ridge, TN\\
    Email: %
    \IEEEauthorrefmark{1}schmidtda@ornl.gov,
    \IEEEauthorrefmark{2}yinj@ornl.gov,
    \IEEEauthorrefmark{3}mathesonma@ornl.gov,
    \IEEEauthorrefmark{4}bronson@ornl.gov,
    \IEEEauthorrefmark{5}shankarm@ornl.gov%
  }
}

\maketitle

\begin{abstract}
The design and construction of high performance computing (HPC) systems relies 
on exhaustive performance analysis and benchmarking. Traditionally this 
activity has been geared exclusively towards simulation scientists, who, 
unsurprisingly, have been the primary customers of HPC for decades. However, 
there is a large and growing volume of data science work that requires these 
large scale resources, and as such the calls for inclusion and investments in 
data for HPC have been increasing. So when designing a next generation HPC 
platform, it is necessary to have HPC-amenable big data analytics benchmarks. In this 
paper, we propose a set of big data analytics benchmarks and sample codes designed for 
testing the capabilities of current and next generation supercomputers. 
\end{abstract}

\begin{IEEEkeywords}
big data; analytics; hpc; benchmarks; R language; MPI;
\end{IEEEkeywords}

\IEEEpeerreviewmaketitle

\section{Introduction}

Many of the state of the art high performance computing (HPC) systems are 
large national investments that serve as critical instruments in the advancement 
of scientific discovery. In the United 
States, the largest of these machines --- which are among the largest in the 
world --- are managed by the Department of Energy. Due to the unique nature of 
these machines, there is a special acquisition process for procuring new ones. 
One such is the CORAL effort~\cite{coral}, a collaboration between Oak Ridge, 
Argonne, and Livermore national laboratories. This involves specifying certain 
performance and power characteristics for a future machine, which vendors 
may then bid to fulfill. One of the components central to this process is the 
evaluation of a comprehensive set of scientific benchmarks.
These benchmarks are generally specified as computational problems relevant to a 
domain science, and are often backed by a community code. Vendors are then 
expected to evaluate and optimize these codes in order to demonstrate the value 
of their proposed hardware in accelerating computational science. This allows a 
vendor to more rigorously demonstrate the performance capabilities and 
characteristics of a proposed machine, since a machine with good performance on 
the benchmark suite should be good for computational scientists.

A new addition to the benchmarking efforts of CORAL acquisitions will include a 
set of big data analytics benchmarks. Throughout this paper, we will describe 
these benchmarks in depth and provide some sample performance numbers on 
existing machines. But first, we must spend some time describing our 
motivations, which may seem peculiar to both data scientists as well as HPC 
programmers. It is generally accepted today that data and analytics are 
important to scientific advancement and discovery, and as a result, a machine 
for computational scientists must also have demonstrable benefit for data 
science. However, data science did not originate from the same philosophical 
space as modern computational science, particularly HPC. For this reason, data 
science in HPC environments can sometimes feel deeply unnatural. So as one might 
expect, this intersection of data science and HPC imposes some unusual 
constraints which may not be readily apparent unless we outline them in 
some detail. In order to maximally appreciate our motivations, we must have a 
discussion about the differences between traditional HPC and data science, and 
how data efforts outside of HPC differ from typical activities within HPC.

This set of benchmarks is, of course, far from the first set of big data 
analytics benchmarks ever created. Indeed, it would be difficult to completely 
summarize the entirety of that space. However, we attempt to briefly summarize 
some of the more well-known efforts on this front. On the HPC front 
specifically, there has been some very informative work in categorizing 
important kernels for big data analytics, and how these dovetail 
with an HPC way of thinking, such as \cite{miniapps} and \cite{fox2015big}. 
However, works such as these are comparatively quite rare as data science in 
general is still somewhat of an outsider in HPC. Although that has been changing 
thanks to excitement surrounding first big data analytics, and more recently, 
deep learning.

Several well-received big data analytics benchmark suites include 
HiBench~\cite{hibench}, BigBench~\cite{bigbench}, and 
Bigdatabench~\cite{bigdatabench}. However, these are largely married to 
MapReduce its compatriots. Perhaps because of this, many of operations of 
interest measured by these benchmarks are more I/O based, focusing on file and 
database operations (some going so far as to benchmark a ``grep'' equivalent). 
In fact, the use of databases in HPC is actually quite rare. This may in part 
explain why high level SQL-like primitives, which are popular in the data space, 
never managed to gain much traction in HPC. In science, often the data is 
simulated at a very large scale. In these cases it is generally not even 
possible to use a database-like construct even if the researcher wanted to. 

More fundamentally, the term ``computational science'' is usually used 
interchangeably with ``simulation science''. But even among empiricists, much 
of the focus in science today is on generating large volumes of data.
That is, scientists (particularly those in HPC) are usually data 
\emph{producers}, while data scientists, statisticians, and the like are 
generally data \emph{consumers}. But producing or consuming, scientists still 
look quite different from their business world counterparts. Unlike with
business data, which may ``hold its value'' indefinitely (thus necessitating 
database-like solutions), many scientists lose all interest in their dataset 
once the paper is published. This is perhaps why many scientists instead choose 
to store their data using frameworks like the Hierarchical Data Format 
(HDF5)~\cite{folk1999hdf5} and NetCDF-4~\cite{rew2006netcdf}, which generally 
have fairly limited support in the data space.

In addition to big data benchmarks, there are also a number of benchmark suites 
specifically for graph problems such as 
LinkBench~\cite{linkbench} and Graphalytics~\cite{graphalytics}. However, we 
treat these kinds of operations as being separate from big data analytics 
benchmarks in our categorization for vendors, and have a dedicated graph 
benchmark. So for the remainder of this paper, we will ignore these entirely.

One analytics benchmark suite that we believe stands out for its rigorous and 
thorough approach is Szilard Pafka's 
benchm-ml~\cite{benchm-ml,pafka2017machine}. In the benchmark description, he 
borrows a quip from George Box about models to quite rightly note that ``all 
benchmarks are wrong, but some are 
useful''. And while this particularly thorough benchmarking effort is valuable, 
it is not 
suitable for this particular task. For one, the data sizes used throughout the 
majority of this benchmark suite are actually quite small, topping out on the 
order of tens of gigabytes. Another issue is the large number of packages 
required, which would put an undue burden on the vendors, who are expected to 
optimize codes for their proposed platform. Indeed, none of the packages used 
in this benchmark are optimized for use in HPC environments.

Another consideration is the software itself. On the one hand, there is a lot 
of attention in the data science space on technologies like 
Spark~\cite{zaharia2016apache}. There has been interest in recent years in 
running Spark on HPC platforms. However, Spark and its various Apache siblings 
were designed for the cloud, not HPC. As such, these frameworks generally 
perform significantly worse than HPC technologies on HPC platforms, with a 
typical performance deficit of more than 
an order of magnitude~\cite{bigdataonhpc,reyes2015big,gittens2016matrix}. 
Beyond Spark, there are many fantastic data science and machine learning 
technologies readily used today by practitioners such as 
Scikit-learn~\cite{pedregosa2011scikit} and caret~\cite{caret}. However, these 
generally do not scale beyond a single node, which is generally a non-starter 
for HPC. Finally, there are behemoth frameworks like 
TensorFlow~\cite{abadi2016tensorflow}. While perhaps largely known for its 
artificial neural network capabilities, it is capable of many techniques, such 
as gradient boosted trees. However, its performance can be quite poor for these 
cases~\cite{tf-gbm}.

Then there is the challenge of even building the software. Tensorflow in 
particular is notorious in HPC circles for the frustrations it causes when 
trying to build it from source. This is because generally in HPC, anything one 
wants to use must be built from source, with no automatic dependency 
resolution, and users do not have root. In an attempt to alleviate this 
problem, there has been some general movement towards adopting container 
strategies, notably Singularity~\cite{kurtzer2017singularity} and 
Shifter~\cite{canon2016shifter} which are based off of Docker but designed 
specifically for HPC. However, even this can prove challenging, and 
these tools are not universally supported.

Finally, any tool selected for a benchmark must be readily available, 
documented, and modifiable by the vendor for their own advantage so that the 
process does not privilege one vendor or technology over another. Most new HPC 
systems today consist of heterogeneous nodes, meaning they have a mix of host 
CPUs and accelerator cards. Current examples of these accelerator cards include 
GPUs and other specialized chips like the Intel Xeon Phi. It would be unfair to 
claim a priori that one solution is necessarily better or worse than another for 
a particular task (indeed, that is up to the vendor to prove!). We must 
therefore make every attempt to remain as ``tech neutral'' as possible in the 
benchmark suite.

All this to say: one generally can not take off-the-shelf solutions used by 
data scientists today and expect them to perform well (or even compile!) on an 
HPC cluster. And many existing big data analytics benchmarks are not adequate 
for our 
comparatively niche task. For all of these reasons, we propose a new set of big 
data benchmarks and sample codes designed for demonstrating the data science 
capabilities of current and next generation supercomputers.

\section{Defining the Benchmarks and Reference Implementations}

\subsection{Background}\label{sec:background}

For CORAL activities, the big data analytics benchmarks are one of about 
24 separate sets of benchmarks which 
vendors are expected to optimize and run. These include everything from very 
low level 
systems performance suites to common scientific applications. So in defining the 
big data analytics benchmarks, we must strike a very delicate balance. We do not 
want to 
overload the vendor or waste their time by throwing in the proverbial ``kitchen 
sink''. At the same time, we need to be sure that the system is well poised to 
deliver capabilities for real data science work.

Ultimately, we chose three techniques to comprise the benchmarks: principal 
components analysis (PCA), k-means clustering, and support vector machine 
(SVM). This gives one dimension reduction technique, one unsupervised 
technique, and one supervised technique, respectively. We also specify 
particular algorithmic constraints (the details of which we will outline in the 
following subsection) for how each of these can be solved. In doing so, this 
also gives one (relatively) computationally simple problem via PCA, which should 
be largely memory bound. SVM by comparison is much more computationally 
complex, and 
should be compute bound. k-means lies somewhere in-between and has much more 
communication than the other two benchmarks. In this way we believe we do a 
reasonably good job of spanning not only the major techniques of interest to 
data scientists, but also the different characteristics of the machine, all 
while maintaining minimality of the benchmark suite. Originally we had planned 
on including more sophisticated techniques, such as gradient boosted trees using 
the popular xgboost framework~\cite{chen2016xgboost}. However in the interest of 
time (both that of the vendors as well as our own), these were ultimately 
dropped. We also deliberately excluded deep learning, as we have a separate 
category of benchmarks exclusively for that task.

For each of these kernels, we supply a ``reference implementation''. This is 
a codebase which we have developed to solve these problems within an HPC 
environment and should achieve reasonably good performance. Any vendor could 
simply take this codebase and merely run it on new hardware to see what 
performance bump is achieved purely for the cost of the machine. However, these 
machines have special characteristics that the reference code is not guaranteed 
to fully utilize. Therefore, it is expected (in fact, desirable) that 
vendors will modify the supplied reference code (or even scrap it entirely for 
a highly optimized custom code) to optimize performance for their proposed 
platform. The degree of allowable modifications is bounded in some ways (for 
example, algorithmically), and unbounded in others (third party libraries used, 
the implementation language, etc.). However, recall that one of our primary 
goals was to avoid privileging one vendor or technology over another. This 
creates a special challenge, in that one technology may be better aligned for 
strong scaling with fewer compute nodes, while another may perform better with 
weak scaling on many compute nodes. We discuss this particular issue in more 
detail in Section~\ref{sec:fom}.

In an attempt to avoid these issues, we do not specify a particular data set 
for use in 
the benchmarks, but instead allow the vendors to randomly generate data with 
certainly minimally invasive distributional requirements (that is, they are 
mixes of several Gaussian distributions). However, this makes it effectively 
impossible to validate any given benchmark run, because the way the data is 
generated could impact non-performance characteristics like classifier 
accuracy. Therefore, we chose to split each benchmark into 2 separate pieces: 
the benchmark which is meant to run at large scale, and a separate validation 
test which is meant to prove that the compute kernel underlying the benchmark 
actually works. For each of the benchmarks, the minimum, average 
(mean), and maximum wall clock run times from the processes (typically MPI 
ranks) will be reported. We request that only the given kernel itself be timed,
as opposed to total run time, to discourage minor advantages like switching out 
of a high level, interpreted language to a low level one. The validation tests 
have specific accuracy requirements which we will detail at length in 
Section~\ref{sec:benchmarks}.

Our benchmarks include a weak scaling component. The reader with an HPC 
background likely needs no introduction to this concept; but as we believe the 
material presented here is valuable not only for them but also for data 
scientists, it is worth briefly elucidating the concept. In a strong scaling 
benchmark, a single global problem size is fixed, and computational resources 
are expanded to see how much faster the problem can be solved from the 
additional computational power. By contrast, in a weak scaling benchmark, the 
\emph{local} problem size is fixed. So as more computational resources are 
added, the global problem grows. In an ideal situation, the (growing) global 
problem can be solved in the \emph{same} amount of time. Each of these 
approaches is valuable and can give useful insight into how a code is 
performing. One particular motivation for examining weak over strong scaling in 
HPC is that for many problems at very large node counts, good strong scaling 
can be very difficult to achieve because of the large communication overhead. 
However, it is also possible for a code that does not scale in the weak sense 
to scale in the strong sense.

Finally, we note that all of the reference code (benchmarks and validation) is 
CPU only, and was not tuned or otherwise prepared for accelerator cards such as 
GPUs.

\subsection{The Benchmark Kernels and Datasets}\label{sec:benchmarks}

Each kernel operates on a double precision data matrix of 250 total columns, 
and some number of rows. The number of rows is up to the vendor, so long as the
total problem size is at least 1024 GB. The choices of 250 rows and 
1024 GB here are somewhat arbitrary; but we believe they are defensible. We 
chose 250 columns because many real data science datasets are of the ``tall and 
skinny'' variety, and so this use case should be well 
handled by any big data analytics technology. There are other common data shapes 
that 
might be of interest, for example ``short and wide''; however, again recall 
that the desire was to construct a minimal set of benchmarks that demonstrate 
the capabilities of the machine without becoming a burden to the vendor. The 
choice of 1024 GB is arguably more arbitrary. Essentially we feel that 
anything under a terabyte is not really ``all that big'' (and these are the big 
data analytics benchmarks, after all). Today it is relatively easy to find 
machines with 
nodes that have RAM in excess of 512 GB, both in supercomputing and increasingly 
in the cloud. When you consider the necessary additional memory storage for 
calculations (or copies if using a high level language), a problem size of 1024 
GB should still require several nodes on most machines.

We require each validation test to run on two nodes and use the same kernel as 
its benchmark counterpart. Each validation test uses the famous iris dataset of 
R. A. Fisher~\cite{fisher1936iris}. The rows of the dataset were randomly 
shuffled and the species variable has been coded to 1 for Setosa, 2 for 
Versicolor, and 3 for Virginica. All other settings, such as how to read or 
distribute the data, we leave up to the vendor. Performance measurements are 
not desired, only proof of validation.

\paragraph{PCA}
We require the first and last of the "standard deviations" from a PCA on a 
distributed matrix to be computed by way of taking the square roots 
of the eigenvalues of the covariance matrix. The "first and last" requirement is 
to forbid the use of approximate methods such as those 
in \cite{halko2011finding}. Using the covariance matrix is mathematically 
equivalent to computing the SVD of the mean-centered input matrix, although it 
is computationally easier to use the covariance matrix. We require that the 
data be random normal from a standard normal distribution (mean 0 and variance 
1). For validation, we 
test the underlying SVD kernel. The validation consists of reading the iris 
dataset, removing the species column, factoring the resulting matrix, and then 
multiplying the factored matrices (from SVD) back together. The mean absolute 
error (average of the difference in absolute value) of the two matrices should 
be computed. The test passes if this value is less than the square root of 
machine epsilon for each type (as specified by IEEE 754).

\paragraph{k-means}
We require the construction of the clustering observation labels (class 
assignments) and centroids using Lloyd's algorithm for 2, 3, and 4 cluster 
centers. The data should consist of rows sampled from one of 3 random normal 
distributions: one with mean 0, one with mean 2, and one with mean 10; each 
should have variance 1. The rows should be drawn at random from these 
distributions. For validation, we use a typical unsupervised ``accuracy''
measurement, rand measure~\cite{rand1971objective}. Using 3 centroids (the 
true value), and 100 starts using seeds 1 to 100, the labels for each 
observation should be computed. These will be compared against 
the true values (from the 'species' label of the dataset) using rand measure. 
We require that the largest among these values should be greater than 75\% to 
be considered successful.

\paragraph{SVM}
We require a linear 2-class SVM to be fit using 500 iterations to 
calculate the feature weights. Specifically, the algorithm which should be 
employed is a hard-margin linear 
SVM that uses Nelder-Mead method to minimize the hinge loss function. The data 
should consist of an intercept term together with rows sampled from one of 2 
random normal distributions: one with mean 0, and one with mean 2; each should 
have variance 1. The rows should be drawn at random from these distributions. 
The SVM is not expected to converge given the number of features and iterations; 
this allows for easier comparisons of benchmarks as data sizes/layouts change.
For validation, we again require an intercept term together with the iris 
dataset, but without the species column. The response is computed from the 
species column, taking 1 for Setosa and -1 otherwise. Using a maximum of 500 
iterations to fit an SVM on the data and response, we request the 
accuracy (number correctly predicted). This should be greater than 80\% in 
order to be considered successful.

\subsection{Figures of Merit}\label{sec:fom}

The final piece of the benchmarks is found in our construction of ``figures of 
merit'' (FOM). Loosely, we define our FOM as size of the input data processed 
per unit of time. Specifically, we require for each kernel that an ensemble of 
jobs is run which is large enough to fill the entirety of the proposed 
machine's nodes. All aforementioned rules imposed for the kernels of course 
still apply. For example, each member of the ensemble should have a problem 
size of at least 1024 GB. So for example, if a vendor proposed a 50 node machine 
which required 5 nodes for a 1024 GB problem, then they could choose to run a 
single 50 node job with a dataset far exceeding 1024 GB, or they could run 10 
ensembles of 5 node jobs, or anything in between. This allows for more 
flexibility among vendors with markedly different hardware configurations and 
characteristics.

\subsection{Reference Implementations}

The reference implementations use R~\cite{rproglang}, which is part 
programming language and part data analysis package. While R is still unpopular 
in HPC, it is extremely popular among data scientists~\cite{bobsoftpop}. In 
fact, the IEEE Spectrum 2017 programming language rankings lists it as the 
number 6 programming language~\cite{ieeespectrum2017}, while the 2018 ranking 
lists it at number 7~\cite{ieeespectrum2018}. This high ranking of R among 
programming languages (and not data analysis packages specifically) is all the 
more remarkable given that R may be an exceptional data analysis package, but 
many consider it a less than stellar programming language.

Each of the three kernels is designed to run multi-node via MPI~\cite{MPI1994}. 
We use the MPI bindings from the pbdMPI package~\cite{pbdMPI}, which is part of 
the pbdR project~\cite{pbdR2012} for large-scale, distributed computing with 
the R language. For the distributed matrix and statistics operations 
specifically, we use the kazaam package~\cite{kazaam}, which directly supports 
each of the kernels PCA, k-means, and SVM. This assumes that the rows of a large 
matrix or dataframe are distributed across MPI ranks, and local linear algebra 
problems are solved via the well-known BLAS and 
LAPACK~\cite{lawson1979basic,Anderson99} libraries. Conceptually, this is 
largely similar to how many other frameworks operate, including the Apache Spark 
library Mllib~\cite{meng2016mllib}. However, there are some critical 
advantages to this implementation over other competitors in this space, mostly 
revolving around the use of MPI instead of MapReduce. 

The full set of scripts and packages used for the remainder of this document is 
available from the CORAL-2 Benchmarks site~\cite{coral2-benchmarks}.

\section{Evaluating the Benchmarks on Current and Next Generation Systems}

\subsection{Baselines and Figures of Merit}

Our baseline runs were on the Oak Ridge Leadership Computing Facility (OLCF) 
machine Titan. 
Titan is a Cray XK7 supercomputer with 18,688 AMD Opteron nodes. Each node has 
16 cores with 32 GB of system RAM, for a total of 299,008 cores and over half a 
TB of system RAM. Each node also has an NVIDIA K20X GPU, with 6 GB of device 
RAM. The nodes are connected with a Gemini interconnect, which uses a 3D torus 
topology. In total, Titan has a theoretical peak performance of 27 petaflops 
(roughly 1.5 teraflops per node).

\begin{figure}
  \includegraphics[width=.475\textwidth]{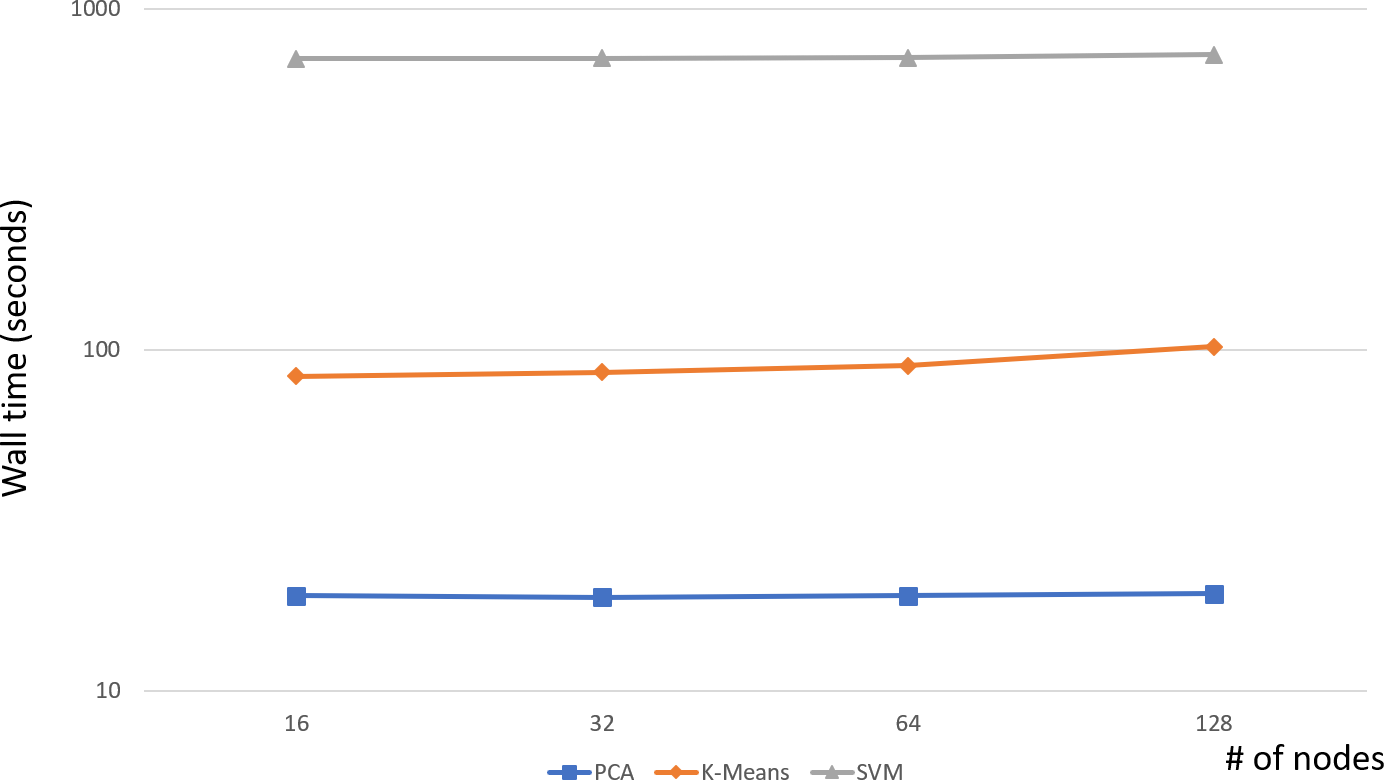}
  \caption{Weak scaling performance (flatter is better) on Titan with the 
reference implementation.}
  \label{fig:titan}
\end{figure}

We found that on Titan, using 8 GB per node was optimal. This left adequate 
room for the additional needs of each benchmark. To meet the 1024 GB 
requirement, this meant that we needed to run on a minimum of 128 nodes. 
However, to get a sense for the weak scaling of the code, we run it at 16, 32, 
64, and 128 nodes. Figure~\ref{fig:titan} shows the results of running the 
benchmarks on Titan at these node counts, with total problem sizes for the 4 
runs of 128, 256, 512, and 1024 GB, respectively. Notice that PCA and SVM are 
almost entirely flat, while k-means has a slight but noticeable uptick to it. 
From this, we expect that we could scale each of these out to many more nodes 
without seeing strong deviations from the flat trend.

\begin{table}
  \centering
  \begin{tabular}{cccc}\hline
    Benchmarks  & k-means       & PCA   & SVM \\ \hline
    $t_a$ (s)   & 83.8          & 24.8  & 631.2 \\
    FOM (TB/s)  & 1.8           & 6.0   & 0.24 \\
  \hline\end{tabular}\label{t1}z
  \caption{Projected Baseline FOM on Titan}
\end{table}
The baseline FOM on Titan is projected as follows. Since the problem size is 
defined to be larger than 1024 GB and the scaling efficiency for 
k-means decreases as the node count increases, we choose the minimum number of 
nodes (128 in our case) for the individual job that can accommodate the input 
data. As a measurement of throughput, we simultaneously run 30 such jobs, which 
accounts for about 20$\%$ of Titan's total node capacity. Afterwards, the 
average maximum wallclock time $t_a$ are calculated for each of the 3 
benchmarks. The projected baseline FOM in TB/s for each benchmark on Titan is 
calculated as $1.024/t_a *(18688/128)$. The details for the individual 
benchmarks are provided in Table~\ref{t1}.

\subsection{Performance Projections on Summit}\label{sec:summit}

At the time of writing, Titan is ranked number 7 on the Top500 supercomputer 
list~\cite{dongarra1997top500}. And so in some sense, Titan represents the 
current generation of supercomputers. Even so, it is quite old by computing 
standards at nearly 6 years of age at the time of writing. To that end, we 
also ran the benchmarks on two additional, newer OLCF systems: Summit and 
SummitDev.

Summit, which is not yet available for general use, will be the 
United States' new largest
supercomputer. While it will officially come online in January of 2019, it is 
at the time of writing the number 1 machine on the Top500. When fully 
operational, it will have roughly 
4600 IBM nodes. Each node will have two 22-core POWER 9 CPUs 512 GB of system 
RAM, giving it more than 200,000 cores and more than 2 PB of system RAM. 
Additionally, each node will have 6 NVIDIA Volta GPUs connected with NVIDIA's 
high-speed NVLink, for a total of around 27,000 GPUs. The nodes are connected 
with a Mellanox EDR InfiniBand interconnect in a non-blocking fat tree topology. 
Each node is anticipated to have a theoretical peak performance in excess of 40 
teraflops. On the other hand, SummitDev is a transitional machine to allow 
researchers to begin testing and porting codes to a Summit-like architecture. It 
has 36 nodes, each with two IBM POWER8 CPUs and four NVIDIA Pascal GPUs.

\begin{figure}
  \includegraphics[width=.475\textwidth]{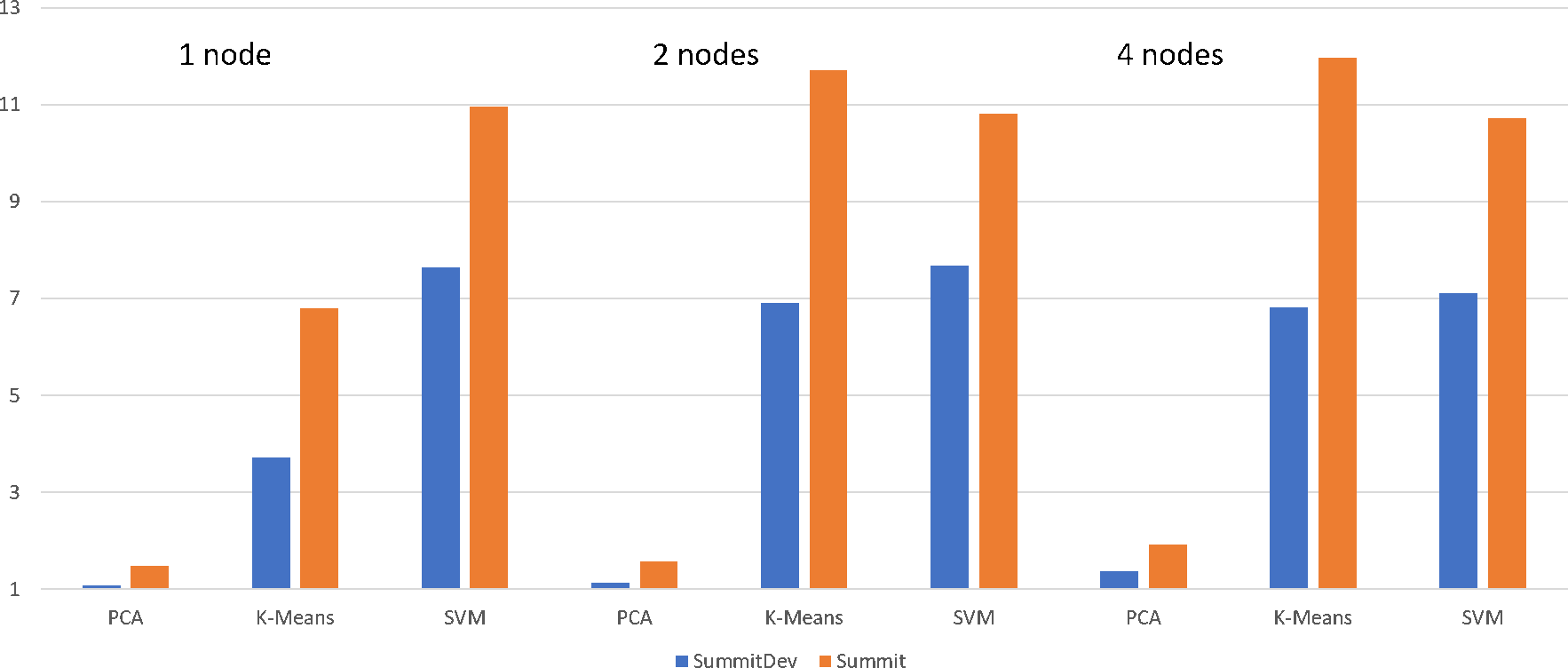}
  \caption{Speedup over the Titan baseline with SummitDev and Summit.}
  \label{fig:speedup}
\end{figure}

For each of the benchmarks, we use 80 MPI ranks per node and 2 threads per rank 
on SummitDev, and 42 MPI ranks per node with 4 threads per rank on Summit for 
the reference implementation, as this was optimal among all configurations we 
tried.

Figure~\ref{fig:speedup} shows the speedup over the Titan baseline on both 
SummitDev and Summit. This performance boost comes without any tuning; it is 
merely the acceleration due to newer, more advanced hardware. The comparisons 
are node-for-node, in that for example, the 2 nodes plot shows the performance 
boost of using 2 nodes of SummitDev or Summit versus using 2 nodes of Titan. 
Each run still uses a problem size of 8 GB per node. We do not scale out to 128 
nodes because in the case of SummitDev, we do not have that many.

Several things are immediately striking about these plots. First, k-means and 
SVM get large 
performance boosts essentially for free, while PCA receives very little 
performance boost. This is relatively unsurprising, given our motivations for 
each benchmark, which we outlined in Section~\ref{sec:background}. Also 
striking is the jump in performance from 1 to 2 nodes for k-means, which is 
sustained from 2 to 4 nodes. This holds true both for comparisons between the 
newer systems and the baseline, but also between SummitDev and Summit. This 
reinforces the suspicion first suggested by Figure~\ref{fig:titan} that the 
k-means benchmark has significantly more network communication than the other 
benchmarks.

Next, we compared the performance of the reference implementation with the 
popular machine learning package H2O~\cite{aiello2016machine}. Like the 
reference implementations, H2O can be run in a multi-threaded/multi-node 
combination. However, H2O is written in Java and a REST API for communication, 
while the reference code's computationally expensive pieces are written in C and 
it uses MPI for communication. Throughout  on both SummitDev and Summit, we use 
80 threads for 1 node and 160 threads per node for the 2 and 4 node case with 
H2O, as this was optimal among all configurations we tried. Additionally, for 
k-means we disabled standardization of input as this was adding significantly 
to its compute time while no such comparable thing was done in the reference 
code.

\begin{figure}
  \includegraphics[width=.475\textwidth]{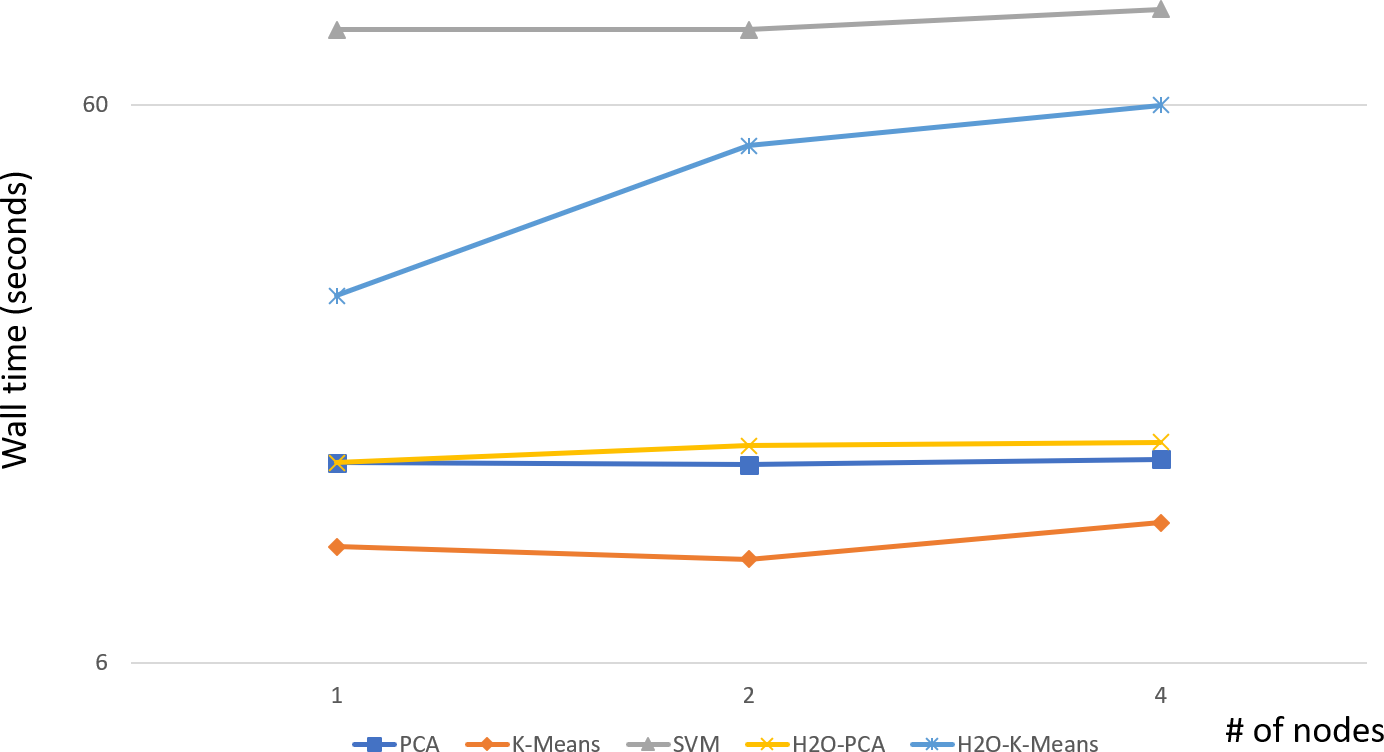}
  \caption{Weak scaling performance (flatter is better) on Summitdev, comparing 
the reference implementation with H2O.}
  \label{fig:sd-weak}
\end{figure}

To compare the two implementations, we run H2O's PCA and k-means kernels (it 
does not have a comparable SVM) with the same 8 GB local problem size on 
SummitDev and examine its weak scaling behavior. Figure~\ref{fig:sd-weak} shows 
the results of this test.

First notice that the two PCA kernels are very close in performance. In fact, 
the two are nearly identical at one node. For 2 and 4 nodes, there is a slight 
divergence, likely due to the advantage of the reference code in using MPI, 
which can better utilize the special interconnect hardware. However, the 
k-means results paint a different picture. The H2O run times are significantly 
higher than those of the reference implementation. We were unable to determine 
why, as the number of iterations for each was fixed and we believe the two to be 
using the same algorithm. It is possible that the H2O implementation is spending 
a lot of effort in attempting to find better initial cluster assignments. This 
would likely reduce the number of iterations needed to convergence, even though 
it makes the initialization more expensive. By comparison, the reference code 
merely uses a single random initialization. Because our number of iterations in 
the benchmark is kept artificially low, it is hard to say what exactly is 
happening here. In the interest of fairness, we believe it is reasonable to give 
the benefit of the doubt here and assume that the extra work is part of a valid 
strategy. So although the absolute numbers may not be directly comparable, the 
relative scaling numbers are. We see the usual slight overall trend upwards in 
the runtimes for the reference implementation. However, there is a very large 
jump from 1 to 2 nodes in the H2O implementation. Again, this is likely due to 
their communication framework not taking advantage of the hardware as MPI can.

\begin{figure}
  \includegraphics[width=.475\textwidth]{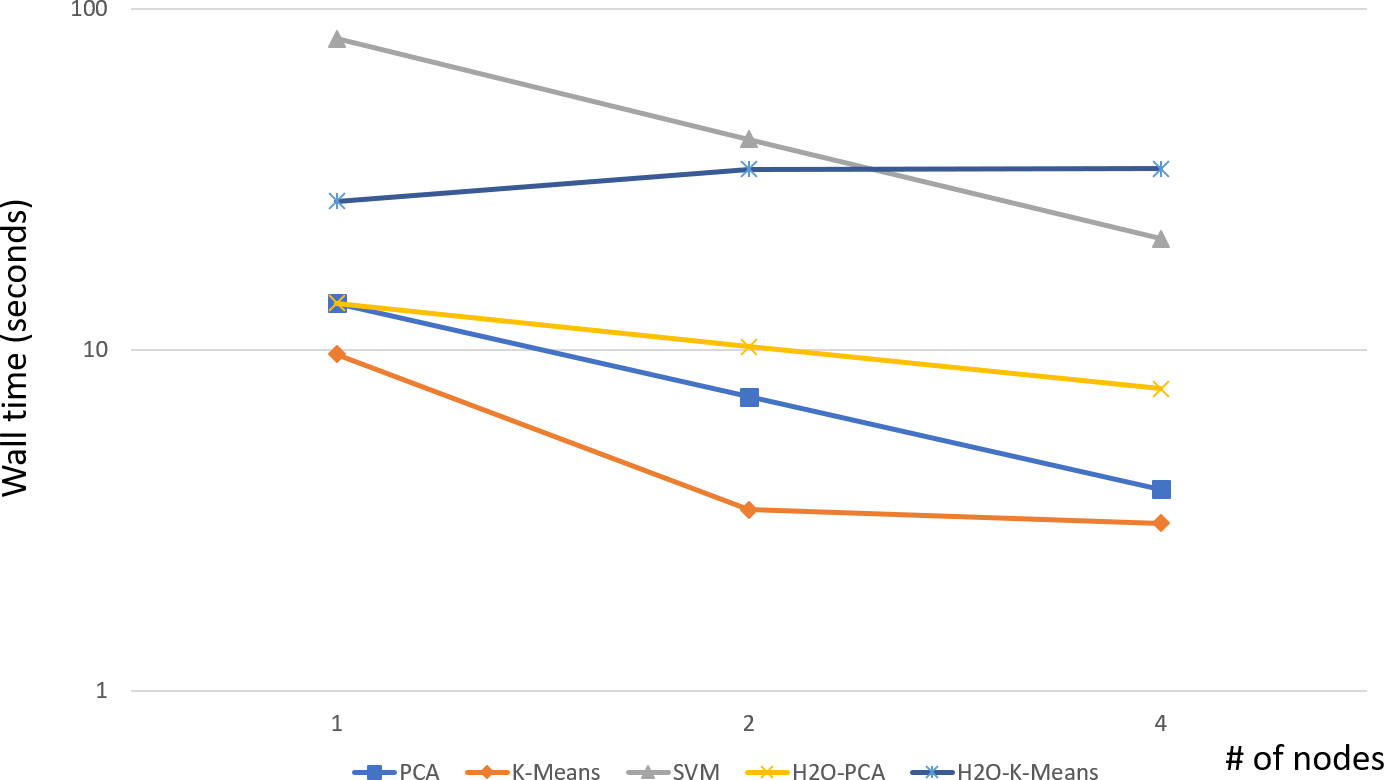}
  \caption{Strong scaling performance (lower is better) on Summitdev, 
comparing the reference implementation with H2O.}
  \label{fig:sd-strong}
\end{figure}

Finally, we made one strong scaling run, the results of which are shown in 
Figure~\ref{fig:sd-strong}. In this case, we fix the problem size across all 
node sizes to 8 GB and we use the same rank and threading layout as before. 
For the reference implementation, we see relatively modest strong scaling from 
each of the kernels at both 2 and 4 nodes. Again with H2O we see what is likely 
larger communication overheads reducing the scalability. In all likelihood, 
both code bases are underperforming in this test due to the small local problem 
sizes at 4 nodes.

It is possible that more tuning would improve the absolute performance of H2O. 
However, the poor scaling to multiple nodes that we see particularly in k-means 
is unlikely to improve because of their current inability to take advantage of 
the machine's advanced network fabrics as MPI does. Essentially, H2O was 
designed for the cloud, not HPC.

\section{Modeling Hardware Performance for Big Data Analytics}
\label{sec:modeling}

Finally, we make several more sets of runs using the benchmark kernels in an 
attempt to understand and model performance of HPC systems for big data 
analytics. To do this, we again employ Summit and SummitDev, but we include two 
additional OLCF systems: Percival and Eos. Percival is a 168 node Cray XC40. 
Each node has a 64 core Intel ``Knights Landing'' (KNL) Xeon Phi processor with 
128 GB of RAM. Eos is a 736 node Cray XC30. Each node has an Intel Xeon 
processor with 64 GB of RAM. Each system has a Cray Aries interconnect with a 
Dragonfly topology.

Throughout, we will consider two problem sizes: 8 GB and 64 GB. While these 
sizes are both relatively small, they are still generally larger than laptop 
sized problems (particularly when one considers the additional memory required 
for workspaces and copies). The smaller of the two runs on a single node of 
each system, and the larger of the two fully saturates at least one node of 
each. For the power systems, we exclusively use one node for each problem size, 
while for the Intel systems we use 2 nodes for Percival and 4 nodes for Eos at 
the 64 GB problem size.

We measure several different potential sources of variation. For one, each 
system has a different architecture: Power 8 (P8) on SummitDev, Power 9 (P9) on 
Summit, KNL on Percival, and Xeon on Eos. We also include the problem size as a 
potential source of variation. Next, we consider different BLAS libraries. On 
each system, we use OpenBLAS~\cite{xianyi2014openblas} as a baseline, and 
compare it against a vendor equivalent, namely MKL~\cite{MKL} on the two Intel 
machines, and ESSL~\cite{essl} on the two IBM machines. Additionally, we use 
two different R versions throughout: 3.4.3 and 3.5.1. There was a major 
internal change in R version 3.5.0 which could affect the performance of 
several of the kernels, k-means in particular. We also try various threading vs 
MPI rank schemas. For the Power machines, we use the same strategy as in 
Section~\ref{sec:summit}, and for the Intel machines, rather than 
oversubscribing we use either 1 MPI rank per core and no multi-threaded BLAS, 
or 1 MPI rank every two cores and 2 BLAS threads. Finally, we treat the 
different benchmark kernels themselves as a source of variation, using them as 
proxies for different workloads (e.g. compute vs memory bound).

We pause here to note that we made numerous attempts to utilize the GPUs on the 
Power machines. Our experiments involved using NVBLAS~\cite{nvblas} in various 
ways. However, we were unable to ever achieve better performance than CPU-only 
configurations using OpenBLAS. We even made several modifications to the 
benchmark kernel codes to try to encourage better performance with the GPUs. For 
example, much of the computation in the SVM code boils down to a fairly large 
matrix-vector product. In the code, this was calling R's matrix product 
function \texttt{\%*\%}, which internally was resolving to the BLAS primitive 
\texttt{DGEMV}. Since \texttt{DGEMV} is not supported by NVBLAS, we modified 
the benchmark kernel to directly call \texttt{DGEMM} so that it would evaluate 
on the GPU. However, the performance results were mixed, and we believe that at 
a minimum, more experimentation is necessary before conclusions can be drawn. 
We know that creating specialized kernels specifically for the GPU will yield 
good performance, but for the purpose of this demonstration, we wished only to 
make relatively minor modifications to the code to get it to run. 
Section~\ref{sec:artifacts} describes our use of NVBLAS in slightly more detail.

Returning to the task at hand, we note explicitly all possible combinations of 
parameters tried. For Percival and Eos, we measure Architecture (KNL and Xeon), 
Problem size (8 GB and 32 GB), BLAS library (OpenBLAS and MKL), R version (3.4.3 
and 3.5.1), Number of threads (1 or 2), and Workload (PCA, KMEANS, and SVM). 
For Summit and SummitDev, we measure Architecture (P8 and P9), Problem size (8 
GB and 64 GB), BLAS library (OpenBLAS and ESSL), R version (3.4.3 and 3.5.1), 
Number of Threads (1, 2, 4, and 8), and Workload (PCA, KMEANS, and SVM). For 
each combination of inputs across both systems, we measure the maximum run time 
and divide it by the problem size for a performance measure in GB/s.

After performing all of the necessary runs, we fit two linear models, one for 
the Power systems and one for the Intel systems. We select an optimal model by 
AIC~\cite{akaike1974new} in a stepwise manner considering only the first order 
terms. The final models are $\log\left(Perf_{Intel}\right)$ is a linear 
combination of $Size$ and $Workload$, while $\log\left(Perf_{Power}\right)$ is 
a linear combination of $Architecture$, $Size$, and $Workload$.

Focusing fist on the Intel model, we note that the final model has an Adjusted 
R-squared of 0.881 with an overall F-statistic of 82.02 (p-value $1.476e-14$). 
The ANOVA table for the model is given in Table~\ref{tab:intelanova}. 
Unsurprisingly, the problem size and workload significant sources of variation. 
The choice of BLAS library and R version being non-significant sources of 
variation are perhaps a little surprising. However, the architecture being 
non-significant is deeply shocking. Evaluating the data, it is not clear 
exactly how this is so, and more study is warranted to understand this behavior.
\begin{table}[ht]
\centering
\begin{tabular}{lrrrr}
  \hline
 & Sum Sq & Df & F value & Pr($>$F) \\
  \hline
  Problem Size & 6.92 & 1 & 36.91 & 0.0000 \\
  Workload & 40.43 & 2 & 107.78 & 0.0000 \\
  Residuals & 5.63 & 30 &  &  \\
   \hline
\end{tabular}
\caption{ANOVA table for the Intel systems.}
\label{tab:intelanova}
\end{table}

For the Power model, we note that the final model has an Adjusted R-squared of
0.954 with an overall F-statistic of 698.1 (p-value $< 2.2e-16$). The ANOVA 
table for the model is given in Table~\ref{tab:poweranova}. As with the Intel 
model, the problem size and workload are, unsurprisingly, statistically 
significant sources of variation. And again, the R version and BLAS library were 
not found to be significant sources of variation. However, as one might expect, 
the system architecture (P8 vs P9) and the number of threads were found to be 
significant.
\begin{table}[ht]
\centering
\begin{tabular}{lrrrr}
  \hline
 & Sum Sq & Df & F value & Pr($>$F) \\
  \hline
  Architecture & 27.68 & 1 & 519.23 & 0.0000 \\
  Workload & 150.08 & 2 & 1407.47 & 0.0000 \\
  Problem Size & 0.67 & 1 & 12.56 & 0.0005 \\
  Threads & 1.06 & 1 & 19.82 & 0.0000 \\
  Residuals & 8.64 & 162 &  &  \\
   \hline
\end{tabular}
\caption{ANOVA table for the Power systems.}
\label{tab:poweranova}
\end{table}

What we can reasonably conclude from these two models is that the node-level 
performance of an HPC system for big data analytics codes like ours is highly 
problem size and workload dependent. CPU architecture may play some role, but 
perhaps not as much as one might expect. Other concerns commonly found at HPC 
centers like choice of boutique BLAS library may add little value, so long as 
one is not using the reference implementation of the BLAS. Trading off MPI 
ranks for BLAS threads or oversubscribing the cores with BLAS threads may lead 
to performance improvements, but this appears to be architecture dependent.

\begin{figure}
  \includegraphics[width=.475\textwidth]{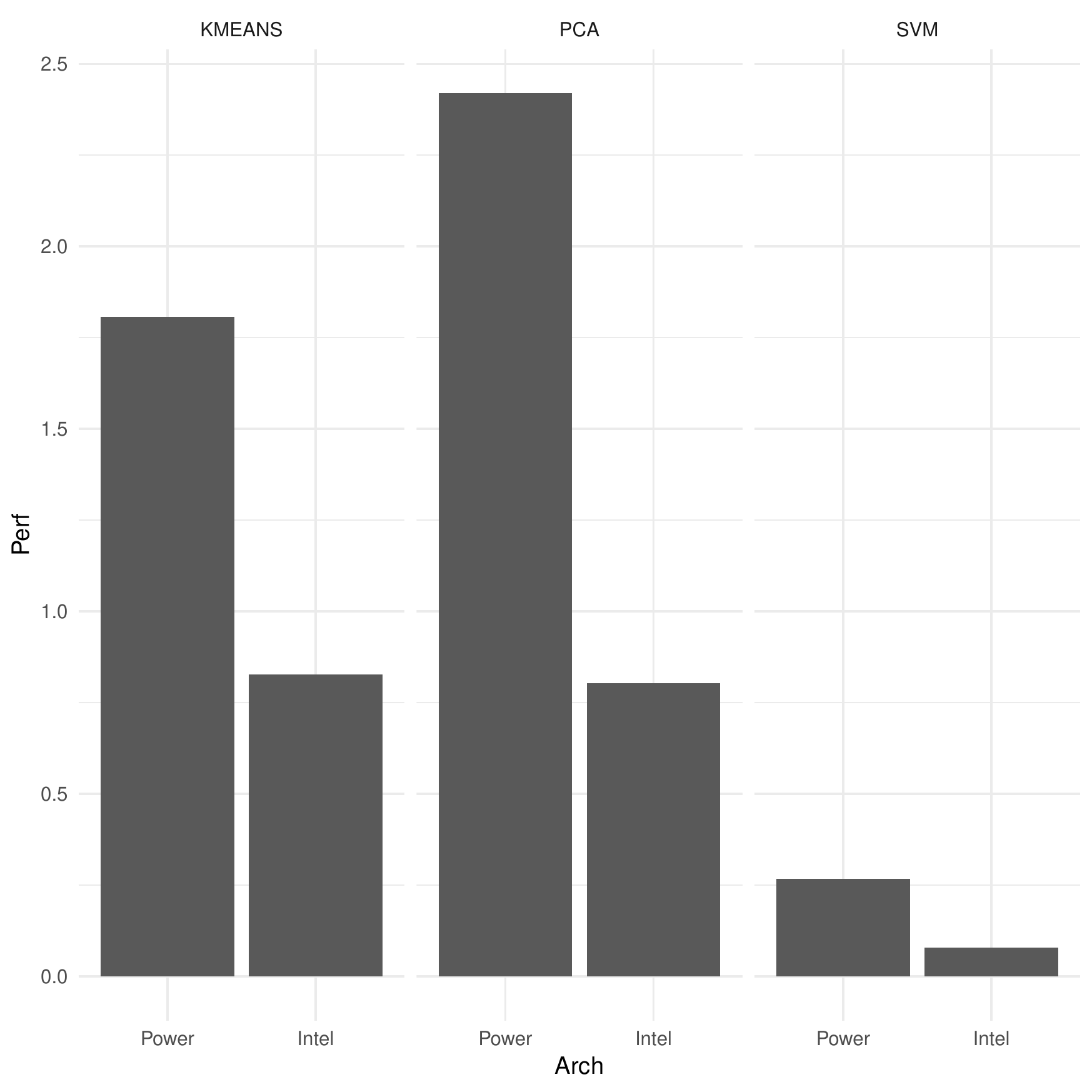}
  \caption{Comparison of average performance across Intel systems 
(Percival/Eos) and Power systems (Summit/SummitDev) on the benchmarks (higher
is better).}
  \label{fig:arches}
\end{figure}

Finally, we average all of the runs per system across groups so that we get a 
single performance number (average GB/s) for each of the three benchmark 
kernels for each of the two systems. Figure~\ref{fig:arches} shows the 
comparison of these values. In each case, the Power systems are quite 
competitive in terms of average performance.
It is worth noting that one key issue this analysis leaves out is cost, of 
both the hardware and power consumption (e.g., flops per watt).  We were unable 
to obtain this information, and it is likely to make this story much more 
complex. However, if price is no concern (and in HPC, that is sadly sometimes 
the case), then the Power systems do appear to be quite fast for big data 
analytics.

\section{Conclusions}

We have described a new set of high performance big data analytics benchmarks 
designed for the peculiarities of HPC systems. We have attempted to motivate 
the necessity for this new work, in light of the large prior work already 
publicly available. In running these benchmarks on existing HPC resources, we 
have seen very good weak scaling and some modest strong scaling at the low node 
counts we used during evaluation. We also used the benchmark kernels to attempt 
to understand and model performance of HPC systems for big data analytics as we 
define it. Invoking Szilard Pafka, we concede that the benchmarks are not 
perfect, but we hope that they are useful.

Finally, we note that the assessment of these benchmarks on Summit is 
preliminary, and therefore the related data and observations reported in this 
paper are subject to change after more thorough studies are performed.

\section*{Acknowledgments}
The views expressed in this paper are those of the authors and do not reflect 
the official policy or position of the Department of the Energy or the U.S. 
Government.

This research used resources of the Oak Ridge Leadership Computing Facility at 
the Oak Ridge National Laboratory, which is supported by the Office of Science 
of the U.S. Department of Energy under Contract No. DE-AC05-00OR22725.

\appendices
\section{Artifact Description}\label{sec:artifacts}

Throughout, we used various versions of the GNU compiler collection, each of 
which is at least version 4.8.

In all cases, we used R version 3.4.3, except in section \ref{sec:modeling} 
where we compare performance between 3.4.3 and 3.5.1. We used pbdMPI version 
0.3-8, although we tested several versions and determined that these different 
versions did not contribute to the performance variation. This was not 
surprising given the changes across recent pbdMPI versions, so we did not bother 
to include it in the final performance analysis above. We used kazaam version 
0.2-0, which included several modifications specifically for these benchmarking 
efforts, as noted in Section~\ref{sec:modeling}. The following configure line 
was used for each build of R:

\begin{lstlisting}
./configure --with-x=no \
  --enable-R-shlib=no \
  --enable-memory-profiling=no 
\end{lstlisting}

For systems libraries, we used OpenBLAS version 0.2.20 across all systems. On 
Percival (Intel KNL), where use of MKL was noted, we used MKL 2018 initial 
release and on Eos (Intel Xeon) we used MKL 2018 update 1. On SummitDev (P8) 
and Summit (P9) where use of ESSL was noted, we used ESSL 5.5.0. For MPI 
libraries, we used Intel MPI on Percival and Eos, with versions corresponding 
to the MKL release versions noted above. On Summit and Summitdev we used OpenMPI 
3.1.0.

For the NVBLAS tests on Summit and SummitDev mentioned in 
Section~\ref{sec:modeling}, we used the following \texttt{nvblas.conf} file:

\begin{lstlisting}
NVBLAS_LOGFILE nvblas.log
NVBLAS_CPU_BLAS_LIB /path/to/libopenblas.so
NVBLAS_GPU_LIST ALL
NVBLAS_TRACE_LOG_ENABLED
#NVBLAS_AUTOPIN_MEM_ENABLED
#NVBLAS_TILE_DIM 2048
\end{lstlisting}
and the \texttt{R\_gpu} script:
\begin{lstlisting}
#!/bin/sh
LD_PRELOAD=/path/to/libnvblas.so 
NVBLAS_CONFIG_FILE=/path/to/nvblas.conf R "$@"
\end{lstlisting}

\bibliographystyle{IEEEtran}
\bibliography{coral2}

\end{document}